\documentclass[3p,times,twocolumn]{elsarticle}

\usepackage{ecrc}

\volume{00}

\firstpage{1}

\journalname{Nuclear Physics B Proceedings Supplement}

\runauth{}

\jid{nuphbp}

\jnltitlelogo{Nuclear Physics B Proceedings Supplement}

\usepackage{amssymb}

\usepackage[figuresright]{rotating}

\begin{document}

\begin{frontmatter}

\title{Diphoton isolation studies}

\author[label1]{Leandro Cieri}
\address[label1]{Dipartimento di Fisica, Universit\`a di Roma ``La Sapienza'' and\\
INFN, Sezione di Roma, I-00185 Rome, Italy}

\author{}

\address{}

\begin{abstract}
We consider the effects of the photon isolation on the production of a pair of photons in hadron collisions.  We study in detail advantages and disadvantages of the standard and `smooth' cone isolation criteria, concerning the theory and the experiment. We put special interest in those kinematic configurations related to recent Higgs boson studies and searches, and finally we show the set of isolation parameters proposed by the Les Houches accord 2013, which serves as a guide to understand the comparison of the theoretical predictions with the data.
\end{abstract}


\end{frontmatter}


\section{Introduction}
\label{sec:Intro}
The production of two isolated photons (diphotons) in high energy hadron collisions is important in order to test the Standard Model (SM) predictions~\cite{Chatrchyan:2011qt,Aaltonen:2011vk,Chatrchyan:2014fsa,Abazov:2013pua,Aad:2011mh,Aad:2012tba}. Besides purely QCD-related considerations, diphoton final states have played a crucial role in the recent discovery of a new boson at the LHC \cite{cha:2012gu,aad:2012gk}, whose properties are compatible with those of the SM Higgs. The diphoton production is also important in many new physics scenarios \cite{:2012afa,:2012mx}, in particular in the search for extra-dimensions~\cite{Aad:2012cy} or supersymmetry~\cite{CMS:2012un}. An improved knowledge of the SM background will help the development of more powerful search strategies and studies for the Higgs boson and for new physics searches.

Describing the production of two isolated photons we have to consider two different contributions: {\em the direct} component which is originated directly from the hard part of the interaction and the fragmentation contribution, in which at least one parton in the final state fragments generating a photon\footnote{If only one parton fragments we refer to this contribution with the name of {\it single fragmentation}. If two partons fragment we call this contribution {\it double fragmentation}.}.

Collider experiments at the Tevatron and the LHC do not perform {\it inclusive} photon measurements. Secondary photons arising from the decays of $\pi^{0}$, $\eta$, etc., overwhelms the signal by several orders of magnitude, and the experimental selection of prompt photons requires {\em isolation} cuts to reject this background. The same isolation cut also affects the prompt-diphoton cross section, which reject the background of secondary photons, in particular by reducing the effect of fragmentation.

The standard cone and the ``smooth'' cone isolation (proposed by Frixione \cite{Frixione:1998jh}) are two of these criteria. The case of the standard cone isolation is easily implemented in experiments, but it only suppresses a fraction of the fragmentation contribution. The smooth cone isolation (formally) eliminates the entire fragmentation contribution, but its experimental implementation (at least in its original form) is complicated\footnote{There is activity in the experimental implementation~\cite{Binoth:2010ra,Blair:1379880,Wielers:2001suas} of the discretized version of the Frixione isolation criterion. An experimental implementation of the smooth isolation criterion was done by the OPAL collaboration~\cite{Abbiendi:2003kf}.} by the finite granularity of the LHC and Tevatron detectors.

The complexity of the calculations can be greatly increased including fragmentation contributions, while the application of appropriate isolation cuts can effectively remove those contributions. How to apply the isolation criteria to the theoretical (TH) tools and how to include the fragmentation contributions consistently by the theory side is one of the subjects of this proceeding.

It is always possible (\textit{i.e.} it has physical meaning) compare theoretical descriptions obtained using the smooth cone isolation criterion and data taken with the standard criterion\footnote{If we use the same isolation parameters for both criteria.}. Moreover, and as we can anticipate, a cross section obtained using the Frixione isolation criterion is always a lower bound for a cross section in which the standard criterion was implemented, and this is valid in the case of the theory and the data. Furthermore, as we show in the next sections, this bound turns out to be an excellent approximation for the cross section calculated with the standard criterion, with an accuracy of the order of the $1 \%$, if {\em tight cuts} are imposed.

Given these results, and the fact that in general it is not possible to exactly match the experimental isolation conditions to the theoretical implementation and viceversa,  we propose a pragmatic accord\footnote{This pragmatic accord is the Les Houches 2013 accord~\cite{Butterworth:2014efa}.} to perform a more precise comparison between the data and the fixed order (or ressummed) calculations, that allows to extend the TH computation for diphoton production up to next-to-next-to-leading-order (NNLO)~\cite{Catani:2011qz} and up to NNLO+NNLL (next-to-next-to-leading-logarithmic accuracy)~\cite{Cieri:2015rqa}; up to NNLO for $Z\gamma-W\gamma$ production or NLO for diphoton production in association with two~\cite{Bern:2014vza,Gehrmann:2013bga,Badger:2013ava} and three~\cite{Badger:2013ava} jets. 

We are interested in Monte Carlo integrators (as  \texttt{DIPHOX}~\cite{Binoth:1999qq}, \texttt{JetFOX}~\cite{Catani:2002ny} or  \texttt{2$\gamma$NNLO}~\cite{Catani:2011qz}, etc) for which the fragmentation component is a purely collinear phenomenon. For Monte Carlo generators (parton--shower Monte Carlo), in which the fragmentation photons are emited off from quarks at non--zero angle during the showering process, we recomend the reference~\cite{AlcarazMaestre:2012vp}.

Beyond the standard and ``smooth'' cone criteria, another way to define direct photons is the so-called ``democratic approach'' \cite{GehrmannDeRidder:1997wx}. In this criterion the photons and QCD partons are treated on the same footing when being clustered into jets. Direct photons are then defined by jets containing a photon which carries a large fraction (typically more than 70\%) of the jet energy. A detailed study of this approach in the context of matrix element to parton shower merging has been performed recently in \cite{Binoth:2010ra}. As in the case of the standard cone isolation criterion, the democratic approach also requires the use of the fragmentation contribution in order to define an infrared safe cross section.

This proceeding is organized as follows. In Sect.~\ref{sec:Isolcriteria} we briefly review the smooth and the standard cone isolation criteria. In Sect.~\ref{sec:thissues} we present numerical results using different isolation criteria
and we comment on their comparison using different implementations of the fragmentation component. In Sect.~\ref{sec:summa} we summarize our results.

\section{Isolation Criteria}
\label{sec:Isolcriteria}
In this section we present the standard and ``smooth'' isolation criteria, its advantages and difficulties concerning their theoretical and experimental implementations.

\subsection{The standard cone isolation criterion}
The standard cone isolation prescription is the criterion used by collider experiments. Schematically it can be described as follows. A photon is said to be isolated if, in a cone of radius $R$ in rapidity and azimuthal angle around the photon direction, the amount of deposited hadronic transverse energy $\sum E_{T}^{had}$ is smaller than some value $E_{T \, max}$ chosen by the experiment:
\begin{eqnarray}\label{standardcriterion}    
&\sum E_{T}^{had} \leq E_{T \, max} \;\;\;\; \nonumber\\
&\mbox{inside} \;\;\;\;      
\left( y - y_{\gamma} \right)^{2} +    
\left(  \phi - \phi_{\gamma} \right)^{2}  \leq R^{2}  \;.   
\end{eqnarray} 
$E_{T \, max}$ can be either a fixed value\footnote{This requirement was typically used at the Tevatron and was motivated by the fact that most of the energy in the isolation cone results from the underlying events (and pile-up), and so is independent of the photon energy~\cite{Binoth:2010ra}.} or a fraction of the transverse momentum of the photon ($p_T^{\gamma}\epsilon$, where typically $0 < \epsilon \leq 1$). 

The theoretical description of the production of a pair of isolated photons is complicated by the occurrence of collinear singularities that appear in the final state, when a photon becomes collinear with a parton. A physical (infrared finite) cross section is only obtained when these singularities are absorbed into the fragmentation functions. Then, as a result, in order to define a theoretically well-defined quantity (if we don't use the Frixione criterion) we have to consider the sum of the direct and fragmentation contributions. Once these two contributions are included, one can isolate the photon using the cuts of Eq. (\ref{standardcriterion}) in an infrared safe way.

In addition, if we consider a narrow cone cut, this case has the undesirable effect of making the theoretical prediction unstable~\cite{Catani:2002ny}. This is due to the restriction of the available phase-space for parton emission.  When the size of the cone used is in the limit of the narrow cone ($R\ll 1$, $R\sim 0.1$), earlier studies reveals potencial problems. This lead to a collinear sensitivity in the form of a fairly large dependence on $\ln(1/R)$, which could make the prediction unreliable\footnote{This could even lead to an unphysical result such as an isolated cross section larger than the inclusive one, thereby violating unitarity.} unless these logarithms were resummed as in Ref.~\cite{Catani:2013oma}, restoring the reliability of the calculation.

\subsection{The Frixione or ``smooth'' isolation criterion}
\label{Sec:Isol_Frix}
The isolation criterion proposed by Frixione~\cite{Frixione:1998jh} (see also Ref.~\cite{Frixione:1999gr,Catani:2000jh}) represents an alternative to the standard isolation prescription. This criterion modifies Eq.~(\ref{standardcriterion}) in the following way
\begin{eqnarray}\label{Eq:Isol_frixcriterion}     
&\sum E_{T}^{had} \leq E_{T \, max}~\chi(r)\;, \;\;\;\;\nonumber\\
&\mbox{inside any} \;\;      
r^{2}=\left( y - y_{\gamma} \right)^{2} +    
\left(  \phi - \phi_{\gamma} \right)^{2}  \leq R^{2}  \;,    
\end{eqnarray}  
with a suitable choice for the function $\chi(r)$. This function has to vanish smoothly when its argument goes to zero ($\chi(r) \rightarrow 0 \;,\; \mbox{if} \;\; r \rightarrow 0\,$), and it has to verify \mbox{$\; 0<\chi(r)< 1$}, if \mbox{$0<r<R\,\,.$} One possible choice is
\begin{equation}
\label{Eq:Isol_chinormal}
\chi(r) = \left( \frac{1-\cos (r)}{1-\cos R} \right)^{n}\;,
\end{equation}
where $n$ is typically chosen as $n=1$. This condition implies that, closer to the photon, less hadronic activity is allowed inside the cone. The cancellation of soft gluon effects takes place as in ordinary infrared-safe cross sections, since no region of the phase space is forbidden. That is the main advantage of this criterion: it eliminates all the fragmentation component in an infrared-safe way. By contrast, it can not be implemented within the usual experimental conditions;
the standard way of implementing isolation in experiments is to use the prescription of Eq.~\ref{Eq:Isol_frixcriterion} with a constant $\chi(r)=1$. In any case, from a purely pragmatic point of view, it has been recently shown~\cite{Butterworth:2014efa} that if the isolation parameters are tight enough (e.g., $E_{T~max} < 6~$GeV, $R=0.4$), the standard and the smooth cone isolation prescription coincide at the $1\%$ level at NLO, which is well within the theoretical uncertainty of our predictions.

Comparing Eqs.~(\ref{standardcriterion}) and (\ref{Eq:Isol_frixcriterion}), it is easy to observe that both criteria coincide at the outer cone ($r=R$, $\chi(R)=1$), and due to the presence of the $\chi(r)$ function, which verifies $0\leq\chi(r)\leq 1$, the smooth cone isolation criterion is always more restrictive than the standard one. This condition directly implies that we expect smaller cross sections when we use the Frixione criterion than when we implement the standard one (for both, theoretically and experimentally), if the same parameters\footnote{\textit{I.e}, the same ($E_{T \, max},R$) or ($\epsilon,R$).} are used in both criteria,
\begin{equation}
\sigma_{Frix}\{R,E_{T~max}\}\leq \sigma_{Stand}\{R,E_{T~max}\}\,\,.
\end{equation}
The smooth behaviour of the $\chi(r)$ function and the continuity property in all its domain (including $r=0$), is the main obstacle to implement the Frixione isolation criterion into the experimental situation. First because of the finite size of the calorimeter cells used to measure the electromagnetic shower. The smooth cone criterion must be applied only beyond a minimum distance of approximately $0.1$ (in $\{ \Delta \eta, \Delta \phi \}$ plane), killing its continuity property up to $r=0$. This condition allows a contribution from fragmentation in the innermost cone and we have to check to which extent the fragmentation component is still suppressed. In addition, the transverse energy in the experimental isolation cone is deposited in discrete cells of finite size. Therefore concerning its experimental implementation, the continuity criterion, initially proposed by Frixione has to be replaced by a discretized version consisting of a finite number of nested cones, together with the collection of corresponding maximal values for the transverse energy allowed inside each of these cones. 

Notice that if we want to apply the discretized version of the smooth cone criterion in the theoretical calculations we have two possible choices. First if we apply the standard cone criterion in the most inner cone, we have to consider also the fragmentation contribution in order to obtain an infrared safe cross section. The problem with this approach is that if we use as radius for the cone $R=0.4$ (tipically used by the experiment), the radius of the most inner cone will be easily near to (or less than) $R\sim 0.1$. And as was reported in~\cite{Catani:2002ny,Catani:2013oma} to get a physical cross section we have to resum to all orders the large logarithmic terms that appear due to the use of a narrow cone ($R\ll 1$, $R\sim 0.1$). The second option is to use the smooth cone criterion in the most inner cone. In this sense we avoid the use of fragmentation contribution and also the problem with the use of the narrow cones, because the Frixione criterion is less sensitive to the large logarithmic terms. In particular the diphoton NLO cross section obtained with the Frixione isolation criterion in this narrow cone limit ($R\ll 1$, $R\sim 0.1$) remains smaller than the inclusive cross section.

\section{Theoretical issues}
\label{sec:thissues}
As it was previously stated, the inclusion of the fragmentation contributions complicates the calculation. And in some cases (as in the diphoton production at NNLO in pQCD) it is not available all the machinery to include it at the desired perturbative level of accuracy.

Basically we can find in the literature two different approaches in order to implement the fragmentation component. These two approaches differ in the way in which they consider the effective order of the fragmentation contribution. The approach of the \texttt{PHOX} family is consider that the fragmentation function has an effective order of $\alpha_{EM}/\alpha_{S}$. The rest of the other approaches consider an effective order of $\alpha_{EM}$ in the fragmentation function. The difference between the two approaches manifests in the final number of Feynman diagrams included in the calculation. Since the \texttt{PHOX} approach consider diagrams which are formally contributions at $\mathcal{O}( \alpha_S^2)$, we refer to this approach as NLO fragmentation\footnote{In this case we are including some diagrams up to $\mathcal{O}( \alpha_S^2)$, due to the effective order of the fragmentation function $\mathcal{O}(\alpha_{EM}/\alpha_{S}$). Including a fragmentation function has the effective impact of ``reduce one order'' in the strong coupling constant for the given contribution.}. In the rest of the approaches, no effective reduction in the order of the strong coupling constant is considered in the fragmentation function. We refer to these approaches as LO fragmentation. The LO fragmentation approach is considered for example in the $\gamma \gamma$ and $W/Z \gamma$ production at NLO in pQCD in \texttt{MCFM} \cite{Campbell:2011bn}, $\gamma \gamma +$ Jet at NLO in pQCD~\cite{Gehrmann:2013aga}, etc. This procedure could introduce inconsistent results in the presence of the standard cone isolation criterion. The following exercise shows how these problems can easely appear.

Let's consider the diphoton production at the LHC ($\sqrt{s}=8$~TeV). First,  we compare the NLO calculation using the fragmentation at NLO with the results in which the fragmentation is considered only at LO. The acceptance criteria in this case require: $p_T^{\rm harder}\geq 40$~GeV and $p_T^{\rm softer}\geq 30$~GeV. The rapidity of both photons is restricted to \mbox{$|y_\gamma| \leq 2.5$} and $100$~GeV$<M_{\gamma \gamma}<160$~GeV. The isolation parameters are set to the values $n=1$ (in the case of the Frixione criterion) and $R=0.4$, and the minimum angular separation between the two photons is $R_{\gamma\gamma}=0.5$. The remaining isolation parameter $E_{T~max}$ (or $\epsilon$) is varied in order to understand the cross section dependence on it. All the cross sections are obtained using the {\small \textsf{CTEQ6M}} set of parton distribution functions.
We used for this analysis the \texttt{DIPHOX} code that includes the full NLO pQCD description. 

\begin{figure}
\begin{center}
\includegraphics[width=0.45\textwidth]{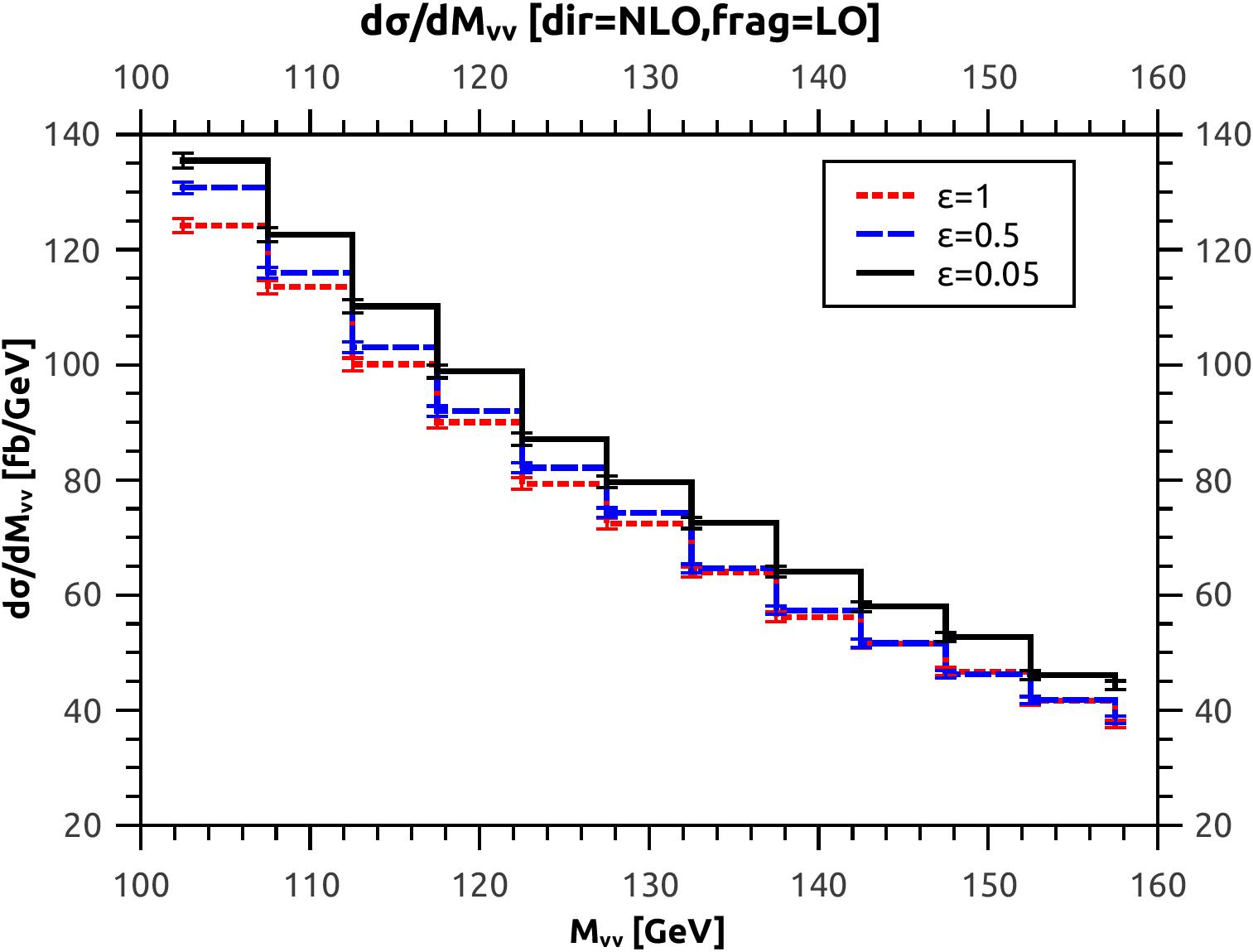}
 \caption{Diphoton cross section as a function of the invariant mass $M_{\gamma\gamma}$ with LO fragmentation component.}
\label{Fig:Isol_1_der}
\end{center}
\end{figure}

\begin{figure}
\begin{center}
\includegraphics[width=0.45\textwidth]{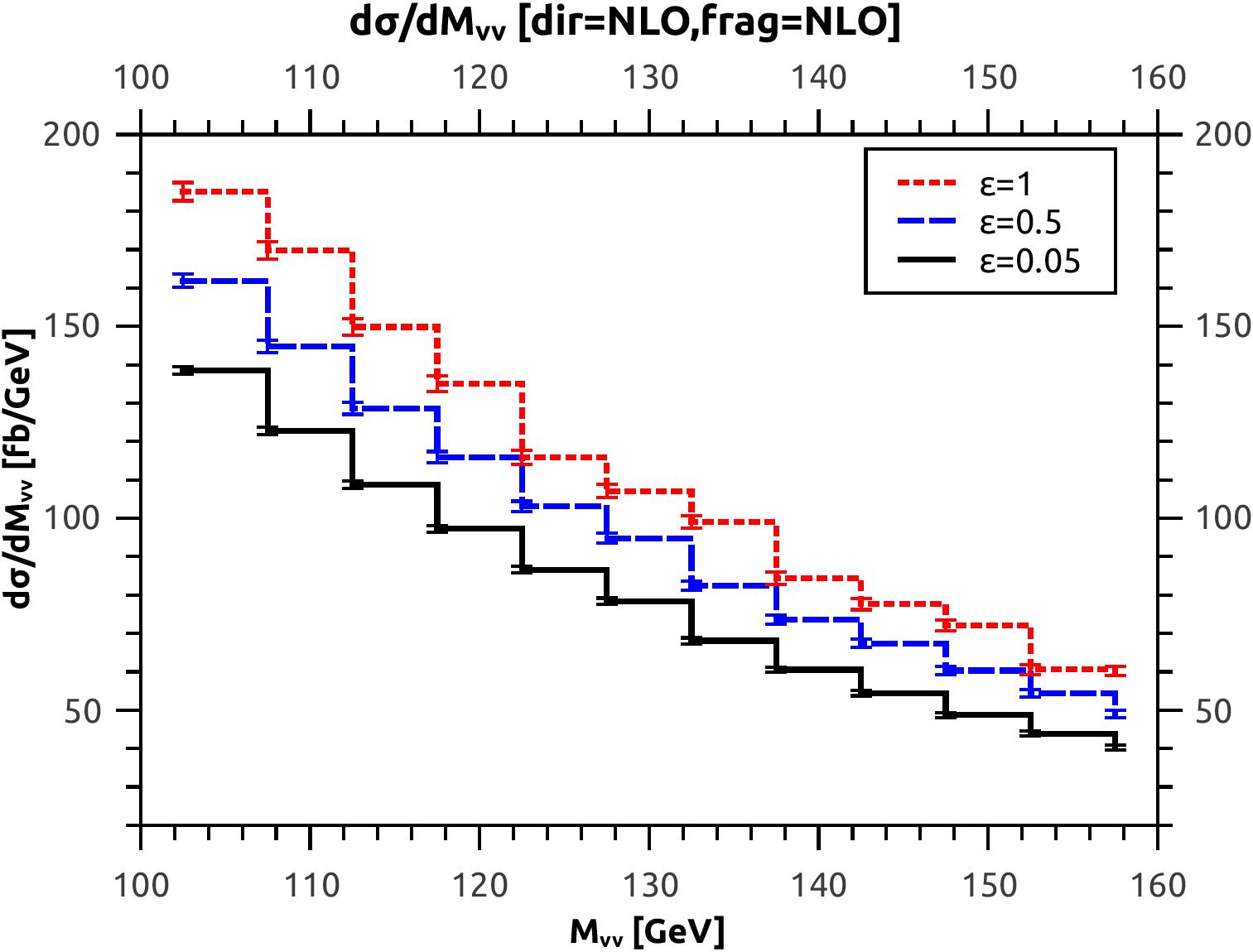}
\caption{Diphoton cross section as a function of the invariant mass $M_{\gamma\gamma}$ with NLO fragmentation component.}
\label{Fig:Isol_1_izq}
\end{center}
\end{figure}

If we use the fragmentation contribution at LO, the results in Fig. \ref{Fig:Isol_1_der} show that we obtain larger cross sections as we impose more severe isolation cuts, which is clearly inconsistent. 
The same behaviour was reported in Ref.~\cite{Gehrmann:2013aga} for $\gamma \gamma +$ Jet at NLO, and we obtained the same for $\gamma \gamma$ production at NLO in pQCD with \texttt{MCFM}. 
On the other hand, the Fig. \ref{Fig:Isol_1_izq} shows the correct behaviour when we consider the full result in which the fragmentation contribution is at the ``same level'' than the direct component (i.e., NLO in this case).

The precedent comparison suggests that one has to be aware that approximating the fragmentation component at one order lower than the direct one can result in unphysical results. The situation can be even more serious when one looks at some extreme kinematical region where the cross section is dominated by higher order contributions.

In Table \ref{Tab:Isol_CMSNLO} we include the results for the corresponding cross section with different isolation prescriptions and parameters. The values presented here help to understand the unexpected behaviour in Fig. \ref{Fig:Isol_1_der}. The analisys contained in this table was made with the cuts implemented by CMS, in a recent measurement of the production cross section for pairs of isolated photons in pp collisions at $\sqrt{s}=7$TeV~\cite{Chatrchyan:2011qt}. We require the harder photon to have a transverse momentum $p_T^{hard}\geq 40$ GeV while for the softer we choose $p_T^{soft}\geq 30$ GeV.
The rapidity of both photons is restricted to $|y_\gamma| \leq 2.5$, without the gap $1.442 \leq y_\gamma \leq 1.556$. Finally, we constraint the invariant mass of the diphotons to the range $100 \,{\rm GeV}\leq M_{\gamma\gamma} \leq 160\,{\rm GeV}$. In the cases in which the standard criterion was applied ($a-f$ in Table \ref{Tab:Isol_CMSNLO}) we observe that as the isolation criterion turns out to be ``loose'', the direct component becomes smaller and the fragmentation component larger. The sum of them behaves as expected with respect to the isolation parameters, since the increase in the fragmentation component overcompensates the decrease of the direct one. In the standard isolation the theoretical separation between direct and fragmentation components is not physical and the results presented in this note correspond to the conventional $\overline{MS}$ subtraction.
On the other hand, if only the LO calculation is used for the fragmentation contributions, for which the next order QCD corrections are quite large (with $K-$factors exceding 2), the mismatch in the perturbative order spoils the compensation between the behaviour of the NLO direct and the LO fragmentation terms. We can observe the resulting effect in the unphysical behaviour presented in Fig. \ref{Fig:Isol_1_der} ( as one considers less restringent isolation parameters).

\begin{table}[ht]
\begin{center}
\resizebox{7.7cm}{!} {
\begin{tabular}{|c|c|c|c|c|c|c|c|c|}
\hline
&Code & $\sum E_{T}^{had} \leq $& $\sigma_{total}^{NLO}$(fb)& $\sigma_{dir}^{NLO}$(fb) & $\sigma_{onef}^{NLO}$(fb)& $\sigma_{twof}^{NLO}$(fb)&Isolation\\
\hline
\hline
a&\texttt{DIPHOX}& $2$ GeV & $3756$ & $3514$& $239$& $2.6$&Standard\\
\hline
b&\texttt{DIPHOX}& $3$ GeV & $3776$ & $3396$& $374$& $6$&Standard\\
\hline
c&\texttt{DIPHOX}& $4$ GeV & $3796$ & $3296$& $488$& $12$&Standard\\
\hline
d&\texttt{DIPHOX}& $5$ GeV & $3825$ & $3201$& $607$& $17$&Standard\\
\hline
e&\texttt{DIPHOX}& $0.05~p_{T}^{\gamma}$ & $3770$ & $3446$& $320$& $4$&Standard\\
\hline
f&\texttt{DIPHOX}& $0.5~p_{T}^{\gamma}$ & $4474$ & $2144$& $2104$& $226$&Standard\\
\hline
g&\texttt{DIPHOX}& \textit{ incl} & $6584$ & $1186$& $3930$& $1468$&none\\
\hline
h&\texttt{2$\gamma$NNLO}& $0.05~p_{T}^{\gamma}~\chi(r)$ & $3768$ & $3768$& $0$& $0$&Smooth\\
\hline
i&\texttt{2$\gamma$NNLO}& $0.5~p_{T}^{\gamma}~\chi(r)$ & $4074$ & $4074$& $0$& $0$&Smooth\\
\hline
j&\texttt{2$\gamma$NNLO}& $2$ GeV~$\chi(r)$ & $3754$ & $3754$& $0$& $0$&Smooth\\
\hline
k&\texttt{2$\gamma$NNLO}& $3$ GeV~$\chi(r)$ & $3776$ & $3776$& $0$& $0$&Smooth\\
\hline
l&\texttt{2$\gamma$NNLO}& $4$ GeV~$\chi(r)$ & $3795$ & $3795$& $0$& $0$&Smooth\\
\hline
m&\texttt{2$\gamma$NNLO}& $5$ GeV~$\chi(r)$ & $3814$ & $3814$& $0$& $0$&Smooth\\
\hline
\hline

\end{tabular}

}
\end{center}
\caption{{\em Cross sections for the $pp\to \gamma\gamma+X$ process at the LHC at NLO.  All these values are at $1\%$ of statistical accuracy level.}}
\label{Tab:Isol_CMSNLO}
\end{table}

Furthermore, from the cases in which the smooth cone criterion was applied ($h-m$ in Table \ref{Tab:Isol_CMSNLO}) we observed that the result in the smooth cone case always provides a lower bound for the one obtained with the standard criterion (as expected) when the same isolation parameters (energy in this case) are used. In the case of smooth cone isolation the (single and double) fragmentation component is identically null.


It is interesting to note, and with one single exception, that the results for the NLO cross sections computed using different isolation precriptions differ by less than $1\%$. The previous result indicates that using the smooth cone prescription for a theoretical calculation (even when the data is analyzed using the standard one) provides an approximation that it is far much better than the one consisting in the standard prescription with a lowest order calculation for the fragmentation component. 
The only case in which one can observe larger differences (of the order of $10\%$) corresponds to the use a very loose isolation, as  for $\sum E_{T}^{had} \leq 0.5~p_{T}^{\gamma}$, where the fragmentation component in the standard case amounts more than half of the total cross section.
\begin{figure}
\begin{center}
\includegraphics[width=0.45\textwidth,height=6.1cm]{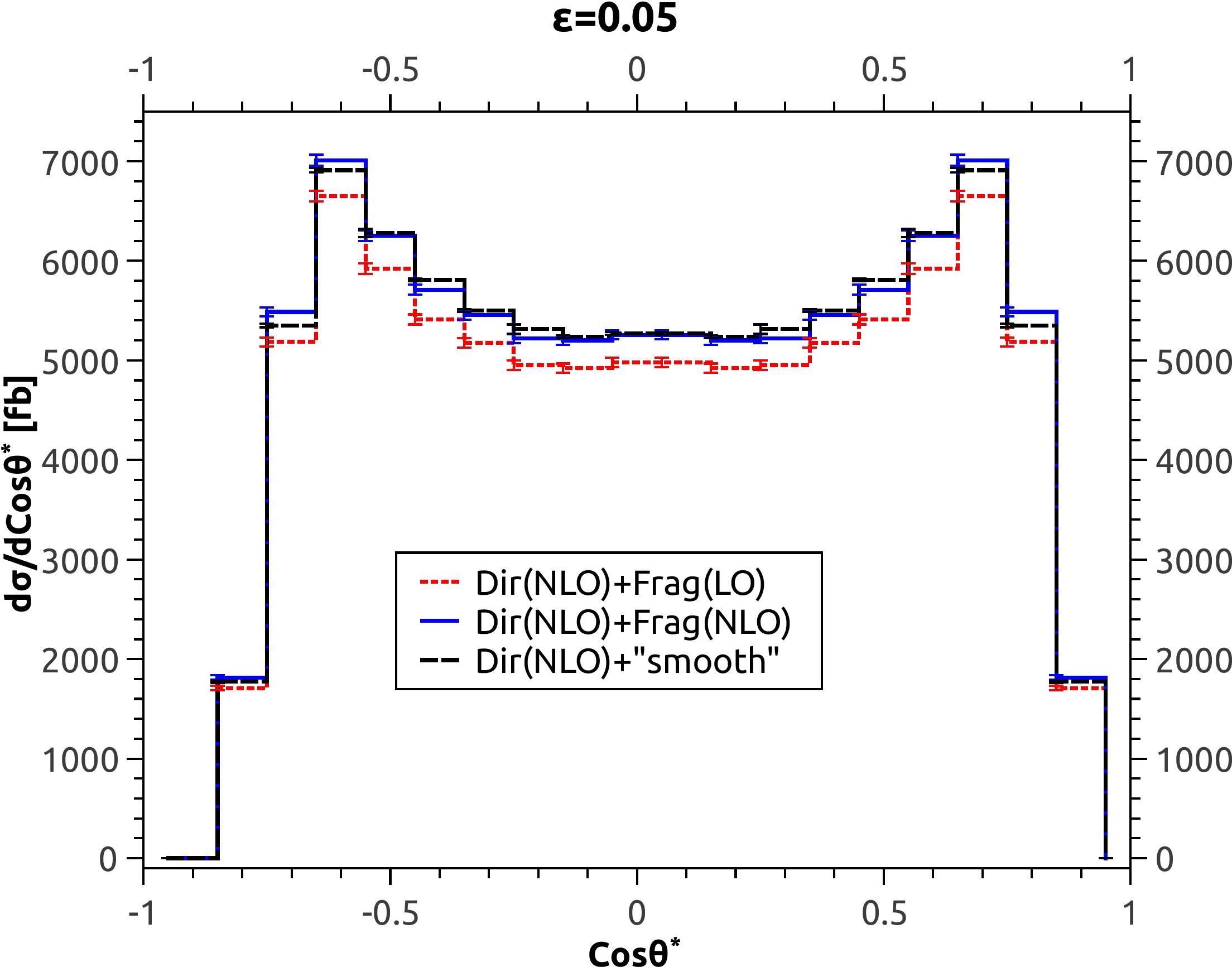}
\caption{Diphoton cross section as a function of the $\cos \theta^{*}$. A comparison between the different isolation criteria is showed. The applied cuts are the same as in Figs. \ref{Fig:Isol_1_izq},\ref{Fig:Isol_1_der}, and are describet in the text.}
\label{Fig:Isol_2_izq}
\end{center}
\end{figure}
\begin{figure}
\begin{center}
\includegraphics[width=0.48\textwidth,height=6.4cm]{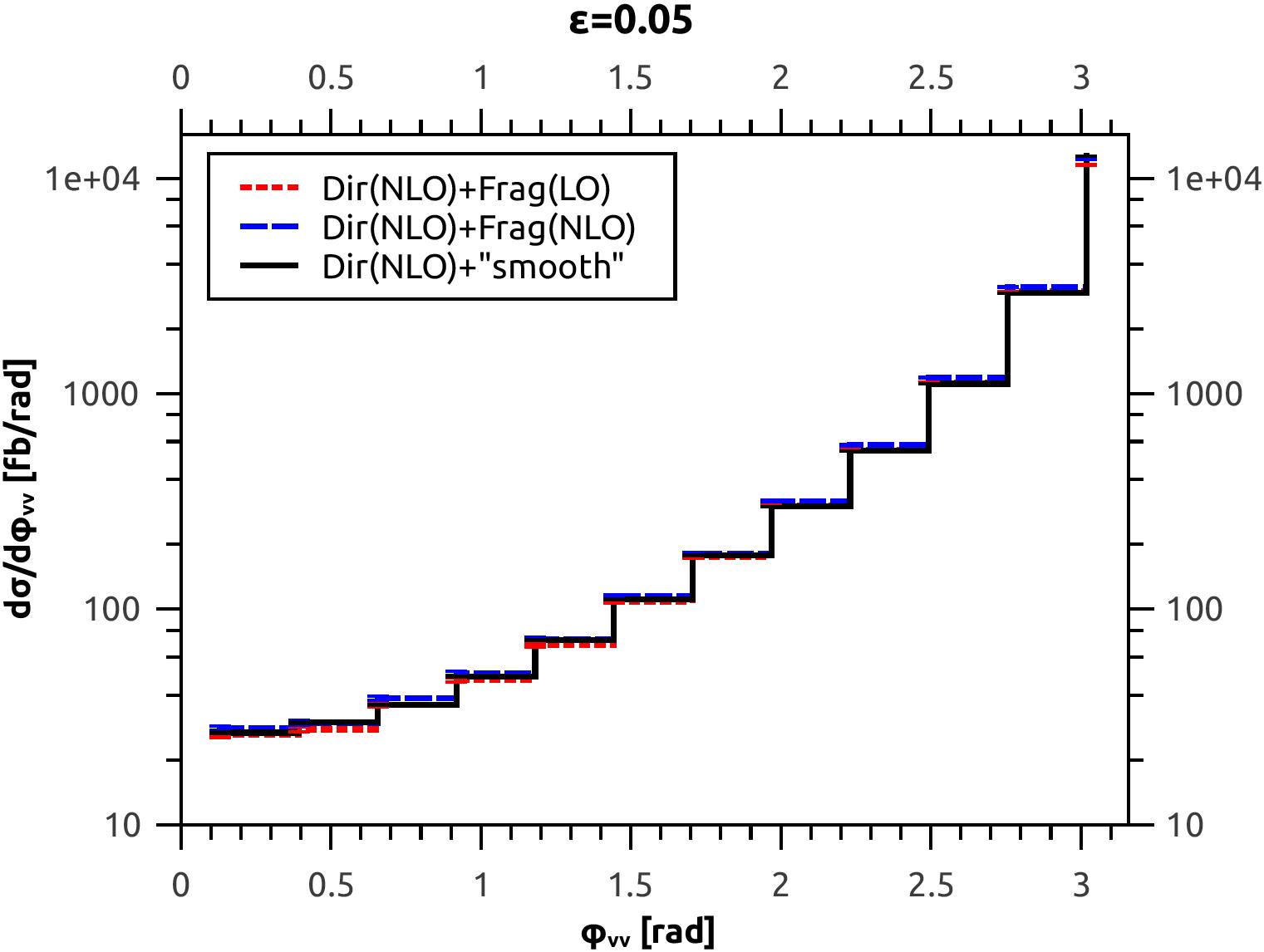}
 \caption{Angular separation between the photons $\Delta \phi_{\gamma \gamma}$ (Right). A comparison between the different isolation criteria is showed. The applied cuts are the same as in Figs. \ref{Fig:Isol_1_izq},\ref{Fig:Isol_1_der}, and are describet in the text.}
\label{Fig:Isol_2_der}
\end{center}
\end{figure}
In all cases we have studied, the smooth cone provides an excellent approximation to the standard result as long as the isolation parameters are tight enough, i.e.  $\sum E_{T}^{had} \leq 0.1~p_{T}^{\gamma}$ or $\sum E_{T}^{had} \leq 5$ GeV for the LHC at 7 TeV. Equivalently, one could define the isolation to be tight enough when the contribution from the fragmentation component does not exceed $\sim 15-20 \%$ of the total cross section.

\begin{figure}
\begin{center}
\includegraphics[width=0.45\textwidth,height=6.4cm]{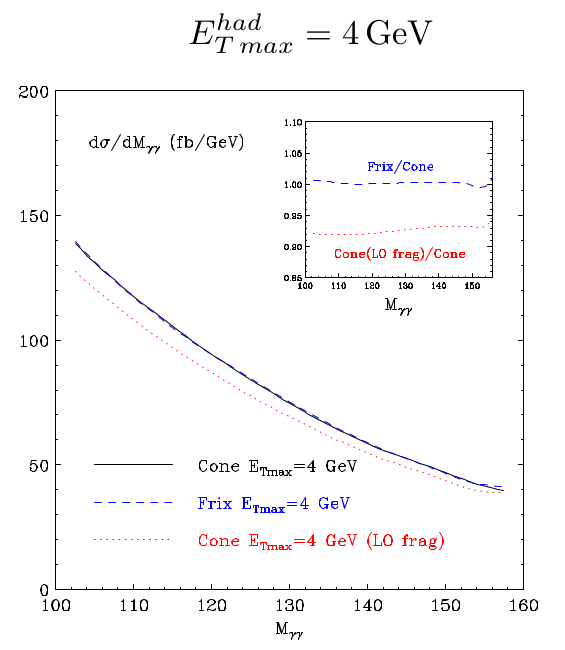}
 \caption{Diphoton cross section as a function of the invariant mass $M_{\gamma\gamma}$.  Cross sections obtained with the standard cone isolation criterion (with LO and NLO fragmentation contributions) are compared with the cross section obtained with the Frixione criterion using the same isolation parameters.
}
\label{Fig:Isol_compToT_der}
\end{center}
\end{figure}

It is known that the effects of fragmentation contributions could be larger in kinematical regions far away from the back-to-back configuration\footnote{The low mass region in the invariant mass distribution, the low $\Delta \phi_{\gamma\gamma}$ distribution and the kinematical regions near to $\cos \theta^{*}=\pm 1$ belong to this case.}, and the approximation could in principle become less accurate for those distributions. In order to check that feature, in Figs. \ref{Fig:Isol_2_izq} and \ref{Fig:Isol_2_der}  we compare the distributions for the full NLO calculation with the standard prescription, the one obtained using only the LO fragmentation component and the result for the smooth cone with $\sum E_{T}^{had} \leq 0.05~p_{T}^{\gamma}$  for $\cos \theta^{*}$ and $\Delta \phi_{\gamma\gamma}$ respectively. In both cases we observe that for all the bins the smooth cone provides the best approximation to the full result, always within a $2.5 \%$ accuracy. A more detailed analysis is presented in Fig. \ref{Fig:Isol_compToT_der},  for the diphoton invariant mass distributions with $\sum E_{T}^{had} \leq 4$ GeV. And we find the same results that in the previous analyses, while using the LO fragmentation component fails to reproduce the full NLO result by up to $6\%$, the smooth cone approximation is always better than $1.5\%$ in the same kinematical region.

\begin{figure}
\begin{center}
\includegraphics[width=0.45\textwidth]{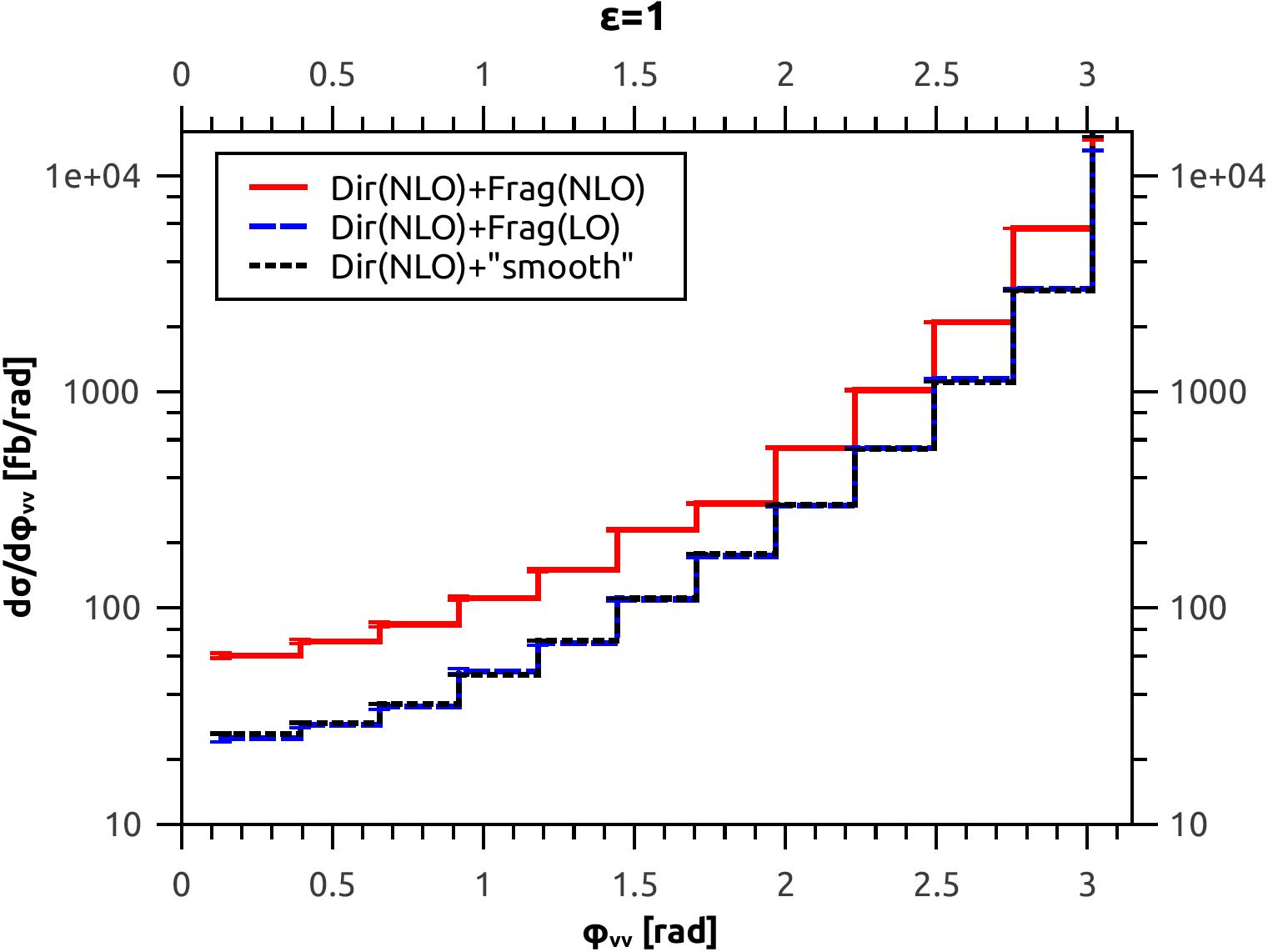}
 \caption{Diphoton cross section as a function of $\Delta \phi_{\gamma\gamma}$. Cross sections obtained with the standard cone isolation criterion (with LO and NLO fragmentation contributions) are compared with the cross section obtained with the Frixione criterion using the same isolation parameters.}
\label{Fig:Isol_3_der}
\end{center}
\end{figure}

In Fig. \ref{Fig:Isol_2_der}, the discrepancies between the full result (or the smooth approach) with the LO fragmentation approximation\footnote{The discrepancies evidently manifest in the invariant mass distribution (see Fig. \ref{Fig:Isol_compToT_der} ) or in kinematical regions far away from $\cos \theta^{*}=\pm 1$ (see Fig. \ref{Fig:Isol_2_izq}).} are ``hidden'' in the $\Delta \phi_{\gamma\gamma}$ distribution, in the bin corresponding to $\Delta \phi_{\gamma\gamma}=\pi$ (the bin containing the back-to-back configurations). Furthermore, in the low $\Delta \phi_{\gamma\gamma}$ region we are dealing with events far away from the back-to-back configuration, and the only configuration that survives at NLO (in these kinematical regions) is the real emission at LO which is for the three cases effectively the same contribution under these conditions. 
\begin{figure}
\begin{center}
\includegraphics[width=0.45\textwidth]{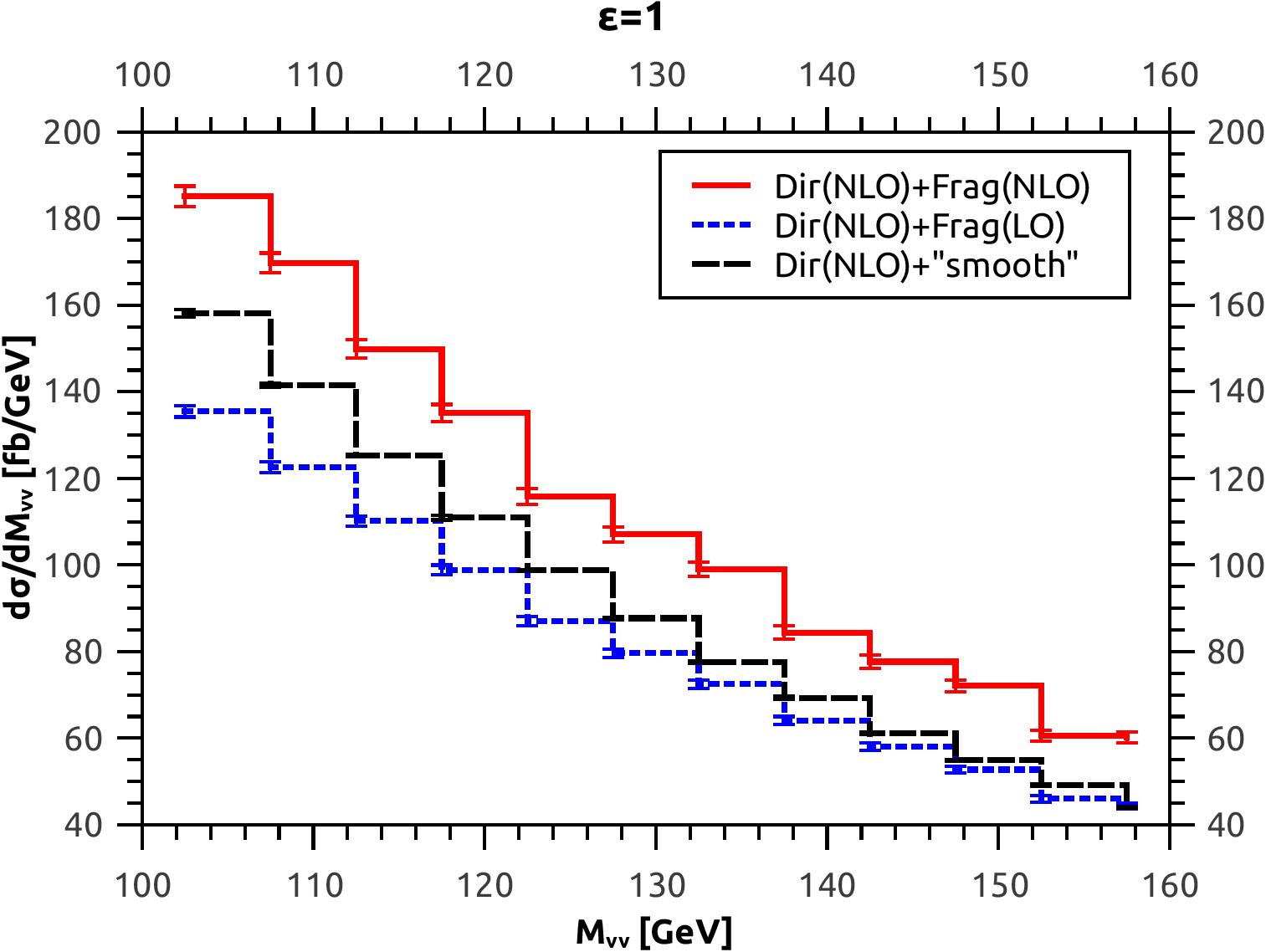}
 \caption{Diphoton cross section as a function of the invariant mass $M_{\gamma\gamma}$. Cross sections obtained with the standard cone isolation criterion (with LO and NLO fragmentation contributions) are compared with the cross section obtained with the Frixione criterion using the same isolation parameters.
}
\label{Fig:Isol_Chir_sinChir}
\end{center}
\end{figure}
If one relaxes the isolation parameters and considers $\epsilon =1$ (which is equivalent to $E_{T}^{max}>40$~GeV, allowing for a huge amount of fragmentation contribution), the effects of fragmentation now strongly manifest at low $\Delta \phi_{\gamma\gamma}$ values (see Fig. \ref{Fig:Isol_3_der}) and in the invariant mass distribution (Fig. \ref{Fig:Isol_Chir_sinChir}). The full result considerably differs from both the LO fragmentation approximation and the smooth cone criterion. We also observe from Fig. \ref{Fig:Isol_3_der}, in the bin corresponding to $\Delta \phi_{\gamma\gamma}=\pi$ (the bin containing the back-to-back configurations) the cross section obtained with the smooth cone criterion provides a better approximation than the LO fragmentation one. 

\section{Summary}
\label{sec:summa}
Considering the results presented in the preceeding sections, it is clear that exists a set of isolation parameters (given a kinematic configuration) for which the standard and smooth cone isolation criteria are in agreement. Matching experimental conditions to theoretical calculations always implies a  certain degree of approximation. Considering the large QCD corrections to processes involving photons (with NNLO essential to understand diphoton data~\cite{Catani:2011qz}) and the agreement (tipically at the $\%$ level for the diphoton case studied here) between the standard and smooth cone TH calculations, the use of the later for TH purposes is well justified.

The \textit{Les Houches 2013}  or "pragmatic" accord that we proposed in Ref.~\cite{Butterworth:2014efa} is the corollary of the previous analysis. Pragmatic in the sense that we do not recommend the experiments to implement the smooth cone isolation, but to proceed to the analysis of the data with the usual standard isolation with cuts tight enough if the interesting observable needs to be an isolated cross section or distribution.
While the definition of "tight enough" might slightly depends on the particular observable (that can always be checked by a lowest order calculation), our analysis shows that at the LHC isolation parameters as $E_{T}^{max}\leq 5$~GeV (or $\epsilon<0.1$), $R\sim 0.4$ and $R_{\gamma\gamma}\sim 0.4$ are safe enough to proceeed.

This procedure would allow to extend available NLO calculations to one order higher (NNLO) for a number of observables, since the direct component is always much simpler to evaluate than the fragmentation part, which identically vanishes under the smooth cone isolation. But it not only concerns NNLO calculations. Theoretical calculations for diphoton production in association with two~\cite{Bern:2014vza,Gehrmann:2013bga,Badger:2013ava} and three~\cite{Badger:2013ava} jets, also apply the smooth cone isolation prescription, because not all the contributions including fragmentation in this case are available.

We also refer to this approach as pragmatic in a numerical sense. The smooth cone isolation applied for the TH calculation is NOT the one used in the experimental data, but considering that NNLO corrections are of the order of $50\%$ for diphoton cross sections~\cite{Catani:2011qz} and a few $100\%$ for some distributions in extreme kinematical configurations, it is far better accepting a few $\%$ error arising from the isolation (less than the size of the expected NNNLO corrections and within any estimate of TH uncertainties!) than neglecting those huge QCD effects towards some "more pure implementation" of the isolation prescription.

\section*{Acknowledgements}
We are very grateful to Stefano Catani and Roberto Bonciani for helpful comments, and to Lisa Collini for her support.

\end{document}